# Observing visible-range photoluminescence in GaAs nanowires modified by laser irradiation

P. A. Alekseev,[1,a)] M. S. Dunaevskiy,[1,2] D. A. Kirilenko,[1] A. N. Smirnov,[1] V. Yu. Davydov[1] and V. L. Berkovits[1]

[1]*Ioffe Institute, Saint-Petersburg, 194021, Russia*

[2]*ITMO University, Saint-Petersburg, 197101, Russia*

We study structural and chemical transformations induced by focused laser beam in GaAs nanowires with axial zinc-blende/wurtzite (ZB/WZ) heterostucture. The experiments are performed using a combination of transmission electron microscopy, energy-dispersive X-ray spectroscopy, Raman scattering, and photoluminescence spectroscopy. For the both components of heterostructure, laser irradiation under atmospheric air is found to produce a double surface layer which is composed of crystalline arsenic and of amorphous $GaO_x$. The latter compound is responsible for appearance of a peak at 1.76 eV in photoluminescence spectra of GaAs nanowires. Under increased laser power density, due to sample heating, evaporation of the surface crystalline arsenic and formation of $β-Ga_2O_3$ nanocrystals proceed on surface of the zinc-blende part of nanowire. The formed nanocrystals reveal a photoluminescence band in visible range of 1.7-2.4 eV. At the same power density for wurtzite part of the nanowire, total amorphization with formation of $β-Ga_2O_3$ nanocrystals occurs. Observed transformation of WZ-GaAs to $β-Ga_2O_3$ nanocrystals presents an available way for creation of axial and radial heterostuctures ZB-GaAs/$β-Ga_2O_3$ for optoelectronic and photonic applications.

## I. INTRODUCTION

GaAs nanowires (NW)s are considered to be promising building elements for new generation of semiconductor optoelectronic and photonic devices,[1] such as solar cells,[2,3] lasers[4] and photodetectors.[5] Firstly, the shape of the nanowires allows to grow semiconductor heterostructures from materials with large lattice mismatch, which cannot be obtained under planar growth conditions.[6-8] Secondly, nanowires can be grown on almost any solid substrate.[9,10] In addition, creation of GaAs NW axial heterostructures whose components have zinc-blende and wurtzite crystal structures is possible.[11] The wide abilities to design GaAs NW heterostructures could be supplemented by their post growth modifications produced by intensive laser irradiation. Focused laser beam is known to be widely used for semiconductor crystal processing. Because of the high surface/volume ratio and of enhancement of optical absorption,[12] the impact of laser irradiation on NWs is anticipated to be more essential. For example, it has been shown recently that focused laser beam can produce in GaAs nanowire a local area of lowered thermal conductivity.[13] For vertically oriented GaAs nanowire it is shown that beam power density higher than $20 Wmm^{-2}$ is sufficient to produce a hollow nanowire morphology.[14] The laser beam also promotes local

---
a) npoxep@gmail.com

oxidation of InAs nanowires and forms an overlayer consisting of indium oxide and crystalline arsenic at nanowire-oxide interface.[15] Recently, a new photoluminescence peak at 1.76eV was observed in lightly doped GaAs nanowires underwent by intensive laser irradiation.[16] It is worth noting in this connection that GaAs nanowires are considered as source of visible light. To obtain optical radiation in visible range, NW axial or radial GaAs-based heterostructures are conventionally created.[17] Due to small size GaAs nanocrystals also allow the obtaining visible optical radiation. Indeed, visible electroluminescence was observed in GaAs nanocrystal stacks in Si nanowires.[18] Thus, investigating the laser-induced modifications of structural, optical and electronic properties of nanowires are highly relevant from scientific and application points of view.

Here we study structural and chemical modifications induced by an intensive laser beam in GaAs NWs and also effect of these modifications on photoluminescence spectra of the nanowires. We clearly show that the photoluminescence peak at 1.76eV arises due to formation of $GaO_x$ amorphous overlayer on the nanowire surfaces under intensive laser irradiation. Laser-induced creation of both radial and axial heterostructures $GaAs/GaO_x$ which exhibit luminescence in the visible range ($\lambda$=500-700nm) is also demonstrated.

## II. EXPERIMENTAL SECTION

The study is performed using a combination of transmission electron microscopy (TEM), energy-dispersive X-ray (EDX) spectroscopy, Raman scattering, and photoluminescence (PL) spectroscopy. Micro-photoluminescence measurements are carried out in the temperature range from 77 K to 300K using Horiba Jobin Yvon T64000 and LabRAM HR spectrometers equipped by a Linkam THMS600 temperature-controlled microscope stage. The measurements are done with continuous-wave (cw) excitation by 532 nm and 325 nm laser lines of Nd: YAG and HeCd lasers, respectively. We use a Mitutoyo 100× NIR (NA=0.90) and a Mitutoyo 50x UV (NA = 0.40) long working-distance objective lenses to focus the incident beam into a spot of ~1 μm diameter, that is enough to measure PL signal from separate nanowire. PL intensity distribution is measured at room temperature using an NT-MDT Ntegra Spectra microspectrometer in confocal luminescence microscopy configuration in reflection mode. Semiconductor laser diode emitting at 473 nm is used for PL pumping. The pumping radiation is focused into a spot of ~1 μm in diameter by lens with numerical aperture NA = 0.95. The excitation density is varied from 1 kW/cm$^2$ to 4 MW/cm$^2$.

Transmission electron microscopy investigations are performed using Jeol JEM-2100F microscope (accelerating voltage 200 kV, point-to-point resolution 0.19 nm) equipped with Oxford Instruments INCA energy-dispersive X-ray spectrometer. As-grown GaAs nanowires are transferred to a conventional TEM copper grid with a lacey carbon film by rubbing the latter over the substrate with the nanowires.

The samples under study are undoped GaAs NWs grown by the vapor-liquid crystal technique under catalytic Ga droplets on GaAs substrates in MBE chamber.[19] The nanowires have diameters and lengths lying in the range of 90-220 nm and of 2-4 microns, respectively. Basically, the NWs have zinc-blende (ZB) crystal structure with stacking faults and twinning, but in vicinity of the top, they have wurtzite (WZ) structure,[20,21] see Fig. 1.

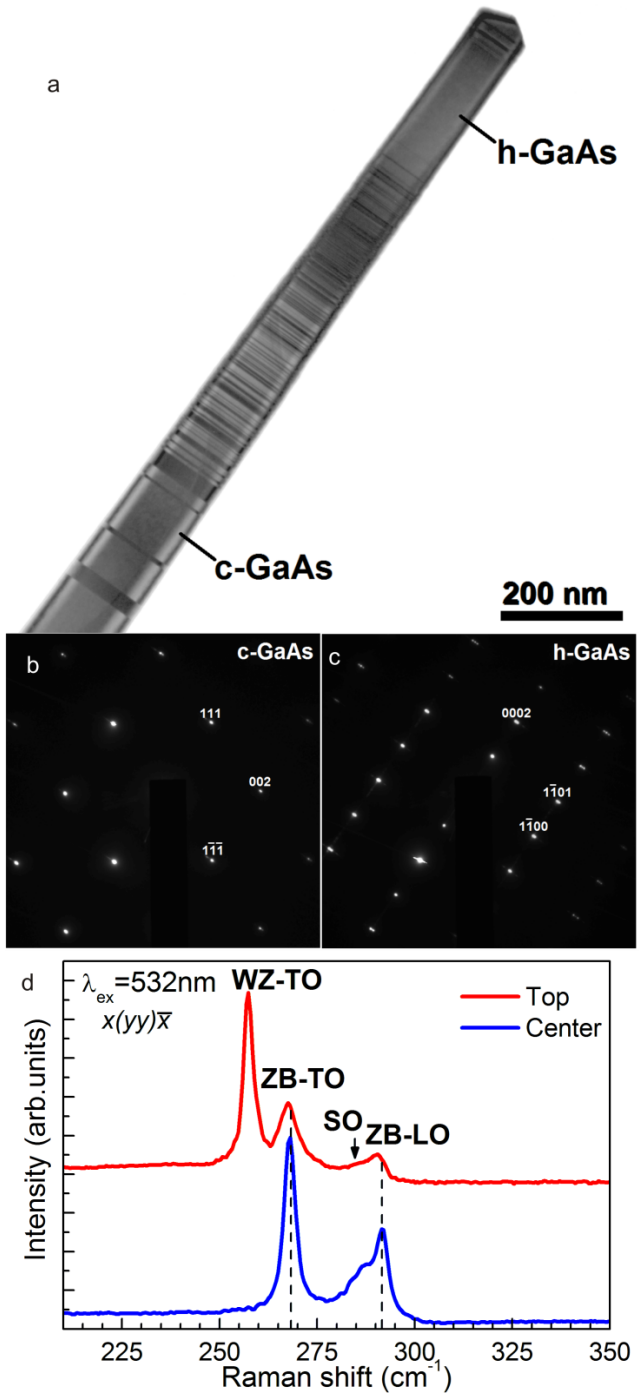

FIG. 1. (a) TEM image of a nanowire with areas with cubic (ZB) and hexagonal (WZ) crystal lattice. The diffraction patterns corresponding to (b) - cubic lattice, (c) – hexagonal lattice (top of the nanowire). (d) Raman spectra measured at the center of the nanowire and at its top.

## III. RESULTS AND DISCUSSION

Figure 2 demonstrates photoluminescence spectra measured from a single GaAs nanowire transferred on Si substrate which is covered by $Si_3N_4$ layer of 10 nm of thickness. The spectra measured at different temperatures (Fig. 2(a)) clearly reveal two peaks lying in 1.4 -1.5 eV and in 1.7-1.8 eV energy intervals. For the first peak, spectral position and its temperature behavior are in a good agreement with known data for the band-edge photoluminescence of GaAs. The second peak at room temperature lies at 1.76 eV, which coincides with the energy interval between spin-orbit-split valence band and conduction band Eg+Δso=1.42+0.34=1.76eV. However, this peak cannot be assigned to such transitions for the following reason. It is known that the value of Δso almost does not vary with temperature.[22] Therefore, if the 1.76 eV peak would originate from optical transitions between spin-orbit-split valence band and conduction band it should shifts with temperature simultaneously with the band-edge PL peak, but that does not occur.

PL spectra measured at room temperature for different beam power densities are shown in Fig. 2(b). As it seen, increase of power density from 100 to 1000 $kW/cm^2$ produces a significant increase of the peak at 1.7-1.8 eV, while the band-edge peak remains almost unchanged. It seems such behavior could be due to structural changes in the NW sample caused by photostimulated reaction proceeding under high density optical irradiation[13].

It is known that the rate of photostimulated reaction in GaAs reduces greatly with temperature lowering.[23,24] To verify this tendency, we measure the sequence of PL spectra shown in Fig. 2(c). Firstly, we measure PL spectra at 79K, then at room temperature, and then again at 79 K. The first spectrum reveals two peaks at energies of 1.48 eV and of 1.50 eV, while no peak is observed in 1.7-1.8 eV range. The peak at 1.48 eV can be assigned to transitions from an impurity band. This band forms due to an unintentional doping, of the nanowire produced by background impurities in MBE chamber. The second peak at 1.50 eV corresponds to band-band transitions in GaAs at T = 79 K. The measured further at room temperature PL spectrum exhibits an additional peak at 1.78 eV which is similar to those shown in Fig. 2(a) and (b). After temperature reduction back to 79 K, the peak at 1.78 eV remains unchanged, but the peak at 1.48 eV disappears. Such findings confirm that at room temperature, high density beam changes irreversibly structure and optical properties of the nanowire. Indeed, it is known that high density optical excitation of GaAs samples promotes to formation of additional non-radiative recombination centers.[25,26] Because a level of the background doping is not high, the arising centers completely suppress radiative recombination from the impurity band. The arising centers also reduce intensity of the band-edge PL. The observed degradation can be obviously explained as induced by photostimulated generation of non-radiative recombination centers.

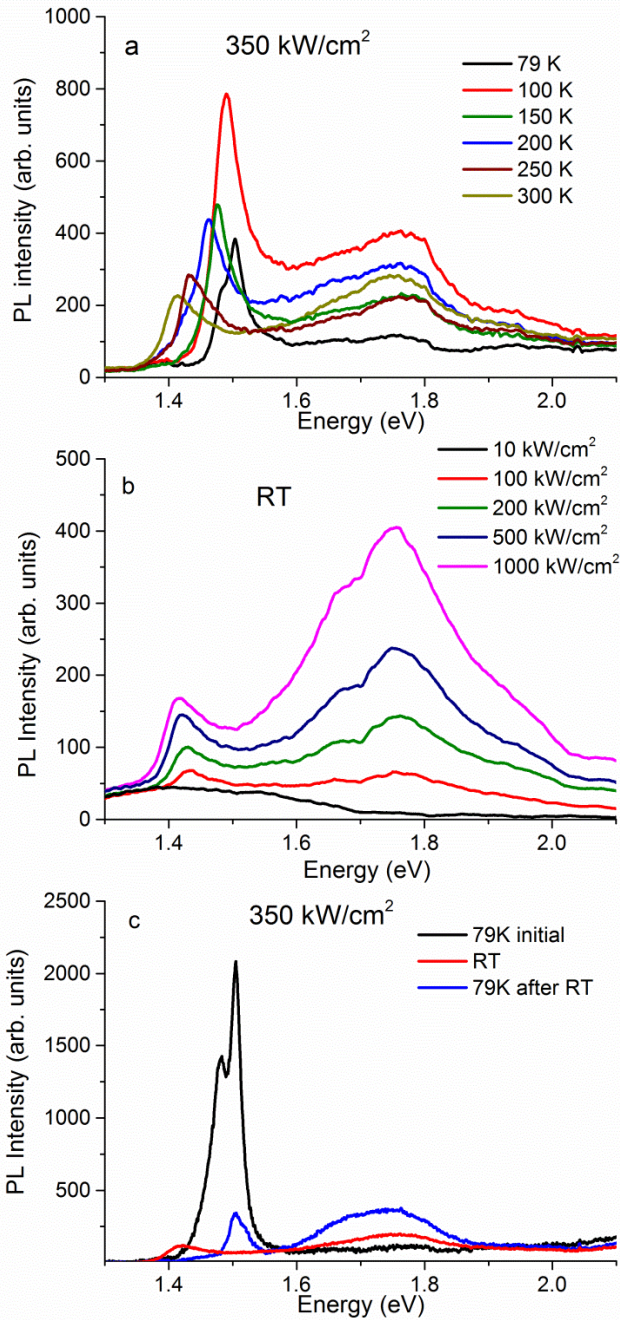

FIG. 2. Sequence of PL spectra measured at the center of GaAs nanowire on the Si/Si$_3$N$_4$ substrate (a) - for beam power density 350 kW/cm$^2$ at 300 K and for further graded temperature decrease down to 79 K; (b) - PL spectra measured at room temperature (RT) for beam power density varying in the interval of (10-1000) kW/cm$^2$; (c) - PL spectra for beam power density 350 kW/cm$^2$ acquired at first at the temperature of 79 K, then at room temperature, then again at T = 79 K.

However, the ideas above do not explain the appearance of peak at 1.76 eV. To identify the nature of this peak, we perform TEM study combined with analysis of the chemical composition of irradiated NWs performed by Energy-dispersive X-ray spectroscopy. The study is performed in the following way. Firstly, the nanowires are transferred to a copper grid for TEM measurements and TEM images are obtained. Then, for selected nanowires photoluminescence spectra at different beam density are measured. Finally, TEM images of the selected nanowires are measured again together with the analysis of their chemical composition. Figure 3 presents TEM images of the same nanowire surface areas before and after 90 second exposition to laser beam with power density of 50 kW/cm² and the corresponding EDX spectra.

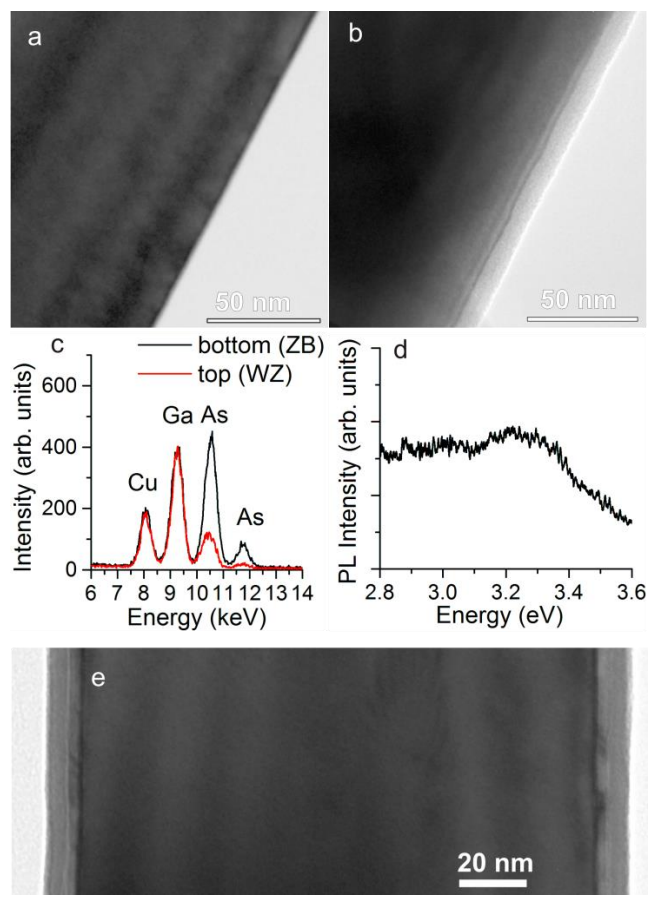

FIG. 3. TEM images of GaAs nanowire surface area (a) before and (b) after exposure to laser beam with power density of 50 kW/cm² for 90 seconds. (c) EDX spectra obtained for irradiated nanowire (black) and for amorphous shell (red). The Cu peak originates from the grid supporting the sample. (d) The PL spectrum obtained after expose to ultraviolet (3.81 eV) laser beam. (e) TEM image of another GaAs nanowire measured after exposure to laser with power density of 50 kW/cm² for 90 seconds.

As it follows from Fig. 3(b), the high density optical radiation produces an amorphous film of 8 nm of thickness on the nanowire surface. Also, in Fig. 3(b) a thin layer between the nanowire surface and the amorphous shell is seen. Formation of the double amorphous surface layer could be related with oxidation of under ambient air conditions. The EDX spectra (Fig. 3(c)) show that the arsenic content in the amorphous shell is 4 times less than in the nanowire body, which indicate that the amorphous shell is enriched with Ga. Recently formation of a crystalline layer of arsenic on InAs nanowire surface after laser exposition was directly observed by TEM and Raman spectroscopy.[15,27] Therefore, one can conclude that the outer shell of GaAs NW is formed mainly from amorphous $GaO_x$ and the interface layer is formed from crystalline As. The processes leading to formation of these layers is discussed below. Figure 3(e) shows TEM image of another GaAs nanowire surface after exposition to laser beam of 50 kW/cm$^2$ for 90 seconds. In this figure, the interface layer between the NW volume and amorphous shell is visible more clearly.

We believe that the amorphous GaO$_x$ layer forming at the nanowire surface is responsible for appearance of the PL peak at 1.76 eV. Indeed, it was shown that amorphous GaO$_x$ shell over crystalline Ga$_2$O$_3$ nanowire is responsible both for "red" luminescence in the 1.7-1.8 eV range,[28] and for UV luminescence in the 3-4 eV range. To confirm our hypothesis, we measure PL spectrum of GaA nanowire using UV (3.81 eV) laser excitation. The obtained spectrum reveals a broad peak at 3.1-3.6 eV shown on Fig. 3(d). Relatively low intensity of the peak is, apparently, due to low density of optical pumping (~100 kW/cm$^2$), and to low transparency of our optical system in the UV spectral range. It should be noted that similar PL peak in 1.7-1.8 eV range is observed in Ga$_2$O$_3$, crystals doped with nitrogen.[29,30] However, in our case, the nitrogen doping was not carried out.

We discuss now how the laser-induced changings of the near surface area in GaAs nanowire depend on the laser power density. Figure 4(a) presents the thickness of the surface GaO$_x$ layer (Fig. S1 in the supplementary material) as a function of laser energy density (the product of the power density and the time). As it seen, with increasing of the power density and duration of exposure, thickness of shell around NWs increases. According to Fig. 4(a), further increase of the energy density can produce the structural changes in almost all entire volume of the nanowire.

Figure 4(b) shows TEM image of GaAs nanowire (ZB/WZ) heterostructure after exposure to laser beam with the energy density of 22.5 MJ/cm$^2$ focused at the point "LF" situated in ZB part of the nanowire. As it seen, the thickness of modified area has maximum value at the point of laser beam focusing.

At the same time, WZ parts of the nanowire situated at the top of NW and in small stack near the base after the light exposure occur to be completely modified. EDX analysis shows that arsenic content in the modified, ex-WZ part near the top of NW is 8 times smaller than that in the center of NW with ZB structure. It is interesting to note that the ex-WZ part of NW contains nanocrystals, see Fig. 4(c). These nanocrystals are also found in the amorphous shell of ZB part of the nanowire. From analysis of diffraction pattern shown in Fig. 4(d), we conclude that the NW area with wurtzite crystal structure under intensive laser irradiation transforms mostly to nanocrystalline phase β-Ga$_2$O$_3$.

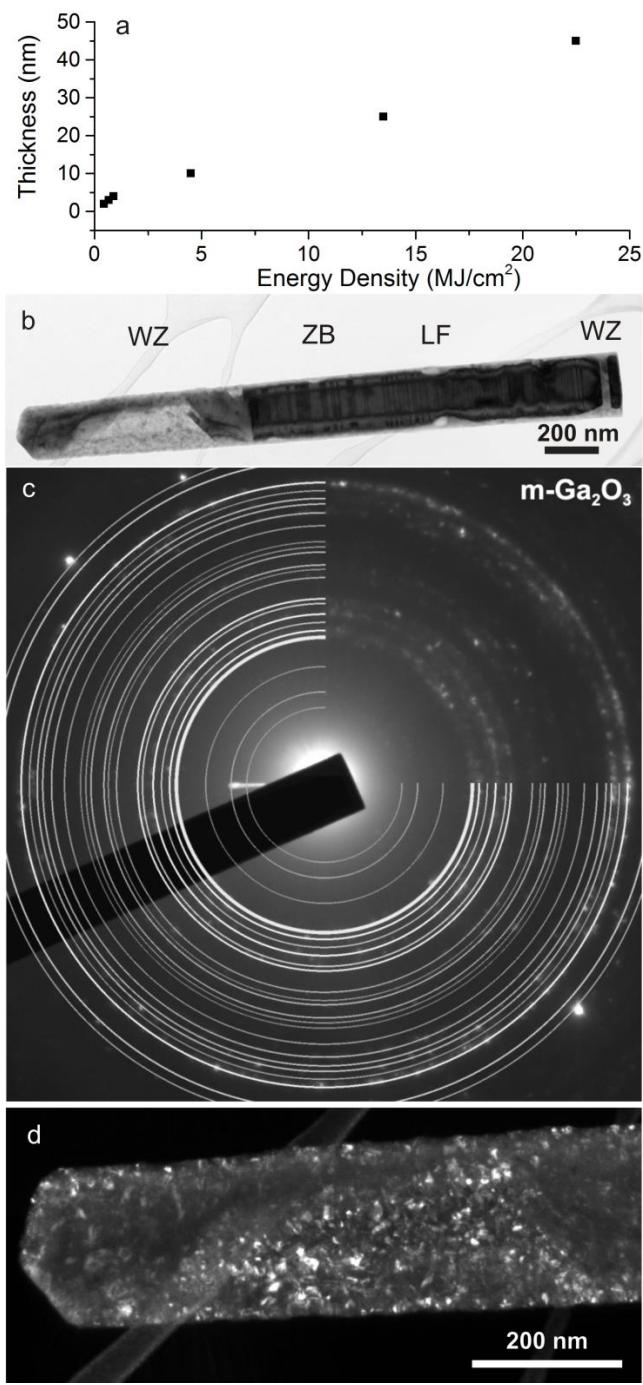

FIG. 4. (a) Dependence of the surface layer thickness versus light energy density. (b) TEM image of the nanowire after laser irradiation with the energy density of 22.5 MJ/cm2 focused at the point "LF". Labels "WZ" and "ZB" indicate NW areas with wurtzite and zinc-blende structures, respectively. (c) Diffraction pattern obtained on NW top with superimposed calculated diffraction rings corresponding to monoclinic Ga2O3 (β phase). (d) TEM dark-field image of the NW top exhibiting nanocrystalline phase.

Nanocrystals $\beta$-$Ga_2O_3$ forms in amorphous shell of ZB - nanowire under increased intensity of laser beam as a result of sublimation of interface As layer. Indeed, no crystalline As interface layer is seen in Fig. 4(b). Figures 5(a) and 5(b) show Raman spectra measured at the top of nanowire (WZ area) and at its center (ZB area) for different beam energy density. The spectra clearly demonstrate that the increase of the power leads to formation of arsenic on the surface that is in agreement with data of references 13-15. In WZ area formation of surface arsenic occurs at a lower beam intensity with respect to ZB area. Further increase of the beam density leads to disappearance of the line corresponding to WZ TO phonons at the top of nanowires, as well as of the line corresponding to arsenic interface layer indicating its evaporation. Analysis of LO phonon position as a function of power density (Fig. S2 in the supplementary material) allows to estimate the temperature of the nanowires.[31] For power density of 1.4 MW/cm$^2$, the nanowire temperature reaches ~ 700 K, that is enough to evaporate the arsenic layer from the nanowire surface.[32]

Now we discuss the effect of compositional and structural changes in nanowires on their PL spectra. Fig. 5(c) shows a map of the PL integrated intensity obtained by laser scanning at a wavelength of 473 nm and for power density of 3 MW/cm$^2$ for a nanowire lying on the Si/Si$_3$N$_4$ substrate. According to Fig. 5(c), the PL intensity distribution is non-uniform, which is caused apparently by inhomogeneity of crystal structure of the nanowire. Figure 5(d) shows the PL spectra recorded at points "1" and "2" indicated in Fig. 5(c). Spectrum obtained at point "1" is the characteristic for the case when laser irradiation removes a layer of arsenic from NW surface and forms the nanocrystals β-Ga$_2$O$_3$. Several peaks can be identified in the 1.7-2.4 eV range in spectrum "1". The appearance of additional peaks is attributed to be due to formation of the nanocrystals β-Ga$_2$O$_3$. It is known that crystalline Ga$_2$O$_3$ reveals a wide PL line in the visible range.[33-35] In spectrum "2" which is characteristic for fully amorphized wurtzite areas with the presence of β-Ga$_2$O$_3$ nanocrystals one can distinguish an additional PL peak at 2.26 eV. This peak can be assigned to photoluminescence of As$_2$O$_3$.[36] Indeed, according to EDX measurement (Fig. S3 in the supplementary material), a small amount of arsenic remains in the wurtzite part of the nanowires after photodegradation.

Because of lower thermodynamic of GaAs with hexagonal structure, nanowire area with wurtzite structure at high density optical excitation becomes completely amorphous. At the same time, area with zinc-blend structure occurs to be covered with the GaO$_x$ layer whose thickness is roughly proportional to beam energy density.

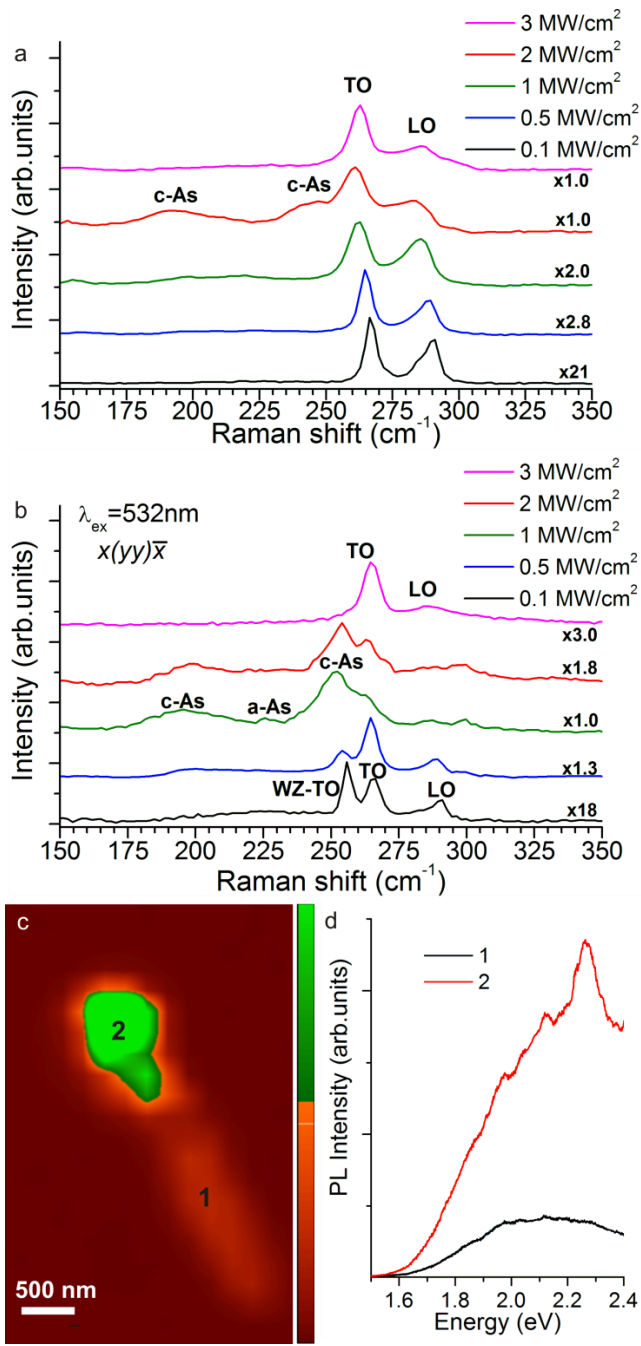

FIG. 5. Raman spectra for different beam power density measured in area with (a) cubic lattice; (b) in area with a hexagonal lattice. (c) Integrated PL intensity distribution for GaAs NW lying on the Si/Si$_3$N$_4$ substrate obtained by laser scanning at power density 3 MW/cm$^2$. (d) PL spectra measured at the points "1" and "2" marked in Fig. 5(c).

It should be noted that formation of the double surface layer crystalline arsenic/amorphous GaO$_x$ is mainly photostimulated reaction, while the removal of As from NW surface is a heating-induced effect. The latter assumption is confirmed by the fact that the removal of arsenic for the nanowire lying on copper grid occurs at optical pumping density which by order of magnitude smaller than that found for the nanowire lying on $Si_3N_4$/Si substrate. This is because thermal transfer is significantly weaker for nanowire lying on copper grid due to a small contact area[14] (Fig. 4(b)).

## IV. CONCLUSIONS

Thus, as a result of high intensity optical irradiation, the surface double layer of crystalline arsenic/amorphous $GaO_x$ forms on the GaAs nanowires. Such layer is responsible for band-edge PL intensity decrease and for emergence of the PL peak at 1.76 eV. Under further increase of optical radiation intensity, all wurtzite part of GaAs NW completely transforms into amorphous $GaO_x$ in which the dominating nanocrystals $β-Ga_2O_3$ give rise visible PL at 2-3 eV range. For zinc-blende part, increase of the beam intensity results in removal of As from surface and in formation of $β-Ga_2O_3$ nanocrystals at the surface.

To conclude, the observed photodegradation of PL intensity in GaAs nanowires under laser irradiation should be necessary taken into account at room temperature luminescence investigations. The account of the photodergradation is also very important for design of GaAs NW-based solar cells using concentrators of solar radiation. In addition, the effect of transformation of GaAs to $GaO_x$ gives an opportunity to create axial and radial heterostructures ZB-GaAs/$GaO_x$ in GaAs nanowires. Indeed, modification of nanowire heterostructure with ZB-GaAs/WZ-GaAs disks by high laser power density should lead to formation of the axial structure ZB-GaAs/amorphous $GaO_x$ with $β-Ga_2O_3$ nanocrystals. Such structures look to be preferable over the ZB-GaAs/WZ-GaAs structures due to large difference in band gap value for GaAs and $GaO_x$. Note, that similar structure may be used to design a single photon source, operating at room temperature. For example, recently developed single photon sources are based on GaAs[37] and InP[38] ZB/WZ superlattices.

**SUPPLEMENTARY MATERIAL**

See supplementary material for the TEM images, EDX spectra of GaAs nanowires and Raman spectra details.

**ACKNOWLEDGMENTS**

The reported study was funded by Russian Foundation for Basic Research (RFBR) according to the Research Project 16-32-60147 mol_a_dk. M.S.D. acknowledge for financial support the Government of Russian Federation (Grant 074-U01). TEM measurements have been carried out using equipment of the Joint Research Center "Material science and characterization in advanced technology". The authors thank Dr. Mihail I. Lepsa (Peter Grünberg Institute, Forschungszentrum Jülich) for providing the GaAs NW samples. The authors also thank Prof. Nikita S. Averkiev, Dr. Ivan Kokurin, Dr. Pavel Petrov, Prof. Victor F. Sapega for fruitful discussion.